\documentclass[%
 aip,
 amsmath,amssymb,
 reprint,%
]{revtex4-1}

\usepackage{graphicx}
\usepackage{dcolumn}
\usepackage{bm}

\usepackage[utf8]{inputenc}
\usepackage[T1]{fontenc}
\usepackage{newtxtext,newtxmath}
\usepackage{microtype}
\usepackage{mathtools}
\usepackage{etoolbox}

\usepackage{braket}
\usepackage{bbm}
\usepackage{siunitx}
\usepackage{multirow}
\usepackage{ifthen}

\usepackage[svgnames]{xcolor}

\definecolor{cset-aps-blueberry}{RGB}{28,128,158}
\definecolor{cset-aps-blue}{RGB}{46,44,184}
\definecolor{cset-aps-turquoise}{RGB}{0,67,88}
\definecolor{cset-aps-limegreen}{RGB}{190,219,67}
\definecolor{cset-aps-green}{RGB}{31,138,112}
\definecolor{cset-aps-yellow}{RGB}{255,225,25}
\definecolor{cset-aps-orange}{RGB}{253,116,0}
\definecolor{cset-aps-red}{RGB}{219,0,43}

\definecolor{sr-trans}{RGB}{207,0,0}
\definecolor{sr-ground}{RGB}{178,202,0}
\definecolor{sr-excited}{RGB}{11,70,135}
\definecolor{sr-mean}{RGB}{0,166,0}
\definecolor{sr-bs}{RGB}{207,0,255}

\usepackage{layouts}

\usepackage{hyperref}
\hypersetup{%
    colorlinks=true,
    linkcolor={cset-aps-red},
    linkbordercolor={cset-aps-red},
    filecolor={cset-aps-orange},
    filebordercolor={cset-aps-orange},
    citecolor={cset-aps-blue},
    citebordercolor={cset-aps-blue},
    urlcolor={cset-aps-green},
    urlbordercolor={cset-aps-green},
    menucolor={cset-aps-limegreen},
    menubordercolor={cset-aps-limegreen},
    breaklinks=true,
    pdfborderstyle={/S/U/W 2},
    pdfpagemode=UseOutlines,
    pdfstartpage={1},
}

\newcommand{\eg}{e.\,g.,}
\newcommand{\ie}{i.\,e.,}
\newcommand{\cf}{cf.}

\newcommand{\ii}{\text{i}}
\newcommand{\dd}{\text{d}}

\newcommand{\DeltaM}{\Delta \mu_0}
\newcommand{\MeanMDM}{\bar{\mu}_\mathrm{DM}}
\newcommand{\DeltaMDM}{\Delta \mu_\mathrm{DM}}

\newcommand{\MeanGDM}{\bar{\gamma}_\mathrm{DM}}
\newcommand{\MeanGEP}{\bar{\gamma}_\mathrm{EP}}
\newcommand{\DeltaGEP}{\Delta \gamma_\mathrm{EP}}

\newcommand{\omDM}{\omega_\varrho}

\newcommand{\SigSin}[2]{
    \ifthenelse{#2=1}%
    {\mathcal{S}(#1)}%
    {\mathcal{S}^{#2}(#1)}
}
\newcommand{\SigCos}[2]{
    \ifthenelse{#2=1}%
    {\mathcal{C}(#1)}%
    {\mathcal{C}^{#2}(#1)}
}

\newcommand{\affTUDa}{Technische Universit{\"a}t Darmstadt, Fachbereich Physik, Institut f{\"u}r Angewandte Physik, Schlossgartenstr. 7, D-64289 Darmstadt, Germany}
\newcommand{\affHAN}{Institut f{\"u}r Quantenoptik, Leibniz Universit{\"a}t Hannover, Welfengarten 1, D-30167 Hannover, Germany}

\newcommand{\orcid}[1]{\href{https://orcid.org/#1}{\includegraphics[width=7pt]{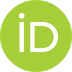}}}

\begin{document}

\title[Clock Transitions Vs. Bragg Diffraction in Dark-matter Detection]{Clock Transitions Versus Bragg Diffraction in\\Atom-interferometric Dark-matter Detection}

\author{Daniel Derr\,\orcid{0000-0002-8690-3897}\,}%
    \email{daniel.derr@physik.tu-darmstadt.de, daniel.derr@gmx.net}
    \affiliation{\affTUDa}

\author{Enno Giese\,\orcid{0000-0002-1126-6352}\,}%
    \affiliation{\affTUDa}
    \affiliation{\affHAN}

\begin{abstract}
Atom interferometers with long baselines are envisioned to complement the ongoing search for dark matter.
They rely on atomic manipulation based on internal (clock) transitions or state-preserving atomic diffraction.
Principally, dark matter can act on the internal as well as the external degrees of freedom to both of which atom interferometers are susceptible.
We therefore study in this contribution the effects of dark matter on the internal atomic structure and the atom's motion.
In particular, we show that the atomic transition frequency depends on the mean coupling and the differential coupling of the involved states to dark matter, scaling with the unperturbed atomic transition frequency and the Compton frequency, respectively.
The differential coupling is only of relevance when internal states change, which makes detectors, \eg{} based on single-photon transitions sensitive to both coupling parameters.
For sensors generated by state-preserving diffraction mechanisms like Bragg diffraction, the mean coupling modifies only the motion of the atom as the dominant contribution.
Finally, we compare both effects observed in terrestrial dark-matter detectors.

\vspace{1mm}
\noindent[This article has been published as part of the \emph{Large Scale Quantum Detectors Special Issue} in\\\href{https://doi.org/10.1116/5.0176666}{AVS Quantum Science \textbf{5}, 044404 (2023)}.]
\end{abstract}

\maketitle

\section{Introduction}
Many envisioned atom-interferometric dark-matter~\cite{Bertone2018} detectors\cite{Abe2021, Badurina2020, Zhan2020, Arduini2023, Buchmueller2023} rely on internal atomic transitions, even though the atom-optical interaction also manipulates the center-of-mass (COM) motion.
While in principle both degrees of freedom can be affected by dark matter (DM), the details of the coupling are key to interpreting and understanding the potential signal measured by atom interferometers (AIs)~\cite{Geraci2016, Figueroa2021, Arvanitaki2018}.
We identify the mean and the differential coupling of the involved atomic states as key quantities and discuss their effect on the atomic transition frequency, as well as on motional effects of terrestrial DM detectors based on atom interferometry.

Terrestrial detectors with both,~\cite{Zhan2020} horizontal,~\cite{Canuel2018, Canuel2020} or vertical~\cite{Badurina2020, El-Neaj2020,Abe2021} orientations are at the planning stage or under construction, for which the site evaluation requires a thorough analysis of the noise environment.~\cite{Savoie2018, canuel2020technologies}
Possible designs differ in their orientations, geometries, source distribution along the baseline, and techniques for atomic manipulation.~\cite{Buchmueller2023}
Here, two-photon transitions~\cite{Hartmann2020} can be used for inducing Bragg diffraction which preserves the internal state and for inducing Raman diffraction which additionally  changes the internal state, although only at hyperfine energy scales.
In contrast, single-photon transitions at optical energy scales can be used to transfer momentum~\cite{Hu2017, Hu2020, Rudolph2020, Bott2023} and to simultaneously generate superpositions of two clock states.~\cite{Ludlow2015}
In differential setups the latter benefits from common-mode suppression of laser-phase noise.~\cite{Yu2011, Graham2013}
However, two-photon transitions are more flexible as they allow to transfer momentum corresponding to optical wavelengths with reasonable laser power without the need for narrow transition lines.

DM may affect both the internal energies of the atom as well as its COM motion.
These effects can in principle be detected by AIs, since they have clock-like properties~\cite{Derevianko2014, Arvanitaki2015, Norcia2017} while being used as accelerometers.~\cite{Graham2016}
The planned detectors will mainly rely on internal transitions, as those have been identified as the dominant contribution of an DM-induced signal.~\cite{Arvanitaki2018, Badurina2022} 
To include DM, one can introduce extensions of the Standard Model that couple to conventional matter,~\cite{Damour1994, Alves2000, Buckley2015, Kimball2023} \ie{} the constituents of the atom, as well as other elementary particles.
By that, each internal energy of the involved atomic states is modified, and through energy-mass equivalence, its motion as well.

In this article we highlight the difference of the internal and external degrees of freedom with the help of two relevant coupling parameters:
Describing the mean coupling of both involved internal states to DM as well as their differential coupling.
Remarkably, both may contribute to the change of the atomic transition frequency and may be detected by AIs based on state-changing diffraction, \eg{} using single-photon transitions.
This is in contrast to most discussions which omit the differential coupling.~\cite{Geraci2016, Arvanitaki2018, Badurina2023, Antypas2022, Filzinger2023, Safronova2018}
Moreover, the relevant energy scales are the unperturbed atomic transition frequency and the Compton frequency, respectively, and are hence of extremely different orders of magnitude.

In AIs based on single-photon transitions like in planned detectors,~\cite{Abe2021, Badurina2020, Zhan2020, Buchmueller2023} the phase from the clock contribution, \ie{} originating from the change of the atomic transition frequency, is dominant.
However, the signal in principle also includes motional effects of the coupling to the COM motion.
In contrast, Bragg-type AIs, as implemented in MIGA,~\cite{Canuel2018} are only susceptible to the latter.
Furthermore, our results are also of relevance for setups that rely on Raman diffraction,~\cite{Zhan2020} where the internal energy scale is in the megahertz regime and much lower than in the envisioned setups built with optical single-photon transitions.
It therefore plays a role in between both limiting cases.
Moreover, the mean coupling identified in our model is closely related to the parameter measured in tests of the Einstein equivalence principle~\cite{Giulini2012, Will2014} (EP), where possible violations of the universality of free fall between different atomic species~\cite{Schlippert2014, Barrett2022} or isotopes~\cite{Asenbaum2020} are studied.
Further, the differential-coupling parameter highlights other facets of the EP, namely violations of the universality of the gravitational redshift~\cite{DiPumpo2021} and of clock rates.~\cite{DiPumpo2023}
To shine light on the influence of these coupling parameters, we furthermore discuss different orders of magnitude of various contributions.

\section{Coupling of Atoms to Dark Matter}
We model DM and violations of the EP by introducing an extension of the Standard Model.
While different approaches are possible that in turn depend on the mass range of interest, our focus lies on ultralight DM.~\cite{Geraci2016}
For that, a classical scalar dilaton field~\cite{Alves2000, Damour2010, Damour2012, Damour1994} $\varrho$ is a simple generic extension.
This dilaton field is linearly coupled~\cite{Damour2010, DiPumpo2021} to all elementary particles and forces of the Standard Model.
Consequently, masses of elementary particles and natural constants become dilaton dependent.
Hence, they also introduce a dependence of composite, bound particles through their constituents.
To describe the resulting effect on atoms, we rely on an effective coupling of their mass and internal states to the dilaton field.

In this article we describe an atom by a two-level system with ground state $\ket{g}$ and excited state $\ket{e}$.
The external degrees of freedom of the atom such as momentum $\hat{p}$ and position $\hat{z}$ are included since the interferometer also acts as an accelerometer.~\cite{Graham2016}
They obey the canonical commutation relation $\left[ \hat{z}, \hat{p} \right] = \ii \hbar \hat{\mathbbm{1}}_\mathrm{ext}$ with the reduced Planck constant $\hbar$.
Therefore, our starting point is the dilaton-modified Hamiltonian
\begin{equation}
\label{eq:start_hamil}
        \hat{H}(\varrho) = \sum_{j=g,e} \bigg[ m_j(\varrho) c^2  + \frac{\hat{p}^2}{2 m_j(\varrho)}+ m_j(\varrho) g(t) \hat{z} \bigg] \otimes \ket{j} \bra{j}.
\end{equation}
Here the relativistic mass defect~\cite{Yudin2018, Sonnleitner2018, Schwartz2019, Martinez-Lahuerta2022, Assmann2023} is incorporated through state-dependent masses $m_j(\varrho)$, which in turn depend on the dilaton field.
In accordance with Einstein's mass-energy equivalence, we find the rest energy $m_j(\varrho) c^2$, where $c$ is the speed of light.
The mass defect already encodes the internal energy difference, which is explicitly given by $[m_e(\varrho)-m_g(\varrho)] c^2$.
Regarding the external degrees of freedom, we consider terrestrial setups modeled by a linear gravitational field.
Both the kinetic and potential energy also become dilaton-dependent.
Additionally, we allow for a time-modulated gravitational acceleration~\cite{Geraci2016} $g(t) = g_0 [1 + \varepsilon_S f(t)]$ with $g_0$ being the gravitational acceleration caused by a source mass and the dimensionless coupling constant $\varepsilon_S$ of the source mass to the dilaton field.
This coupling constant can be defined in principle in analogy to the atomic mass below, \cf{} Eq.~\eqref{eq:mass_expansion}.
Here, $f$ is some time-dependent modulation induced by the coupling of the dilaton field to the source mass.

We assume that the change of the atomic mass due to the dilaton field is small.
Consequently, a first-order expansion of the mass~\cite{DiPumpo2021} $m_j$ at the Standard-Model value $\varrho=0$, \ie{}
\begin{equation}
\label{eq:mass_expansion}
    m_j(\varrho) \cong m_{j,0} \left( 1 + \varepsilon_j \, \varrho \right),
\end{equation}
gives rise to the effective coupling to the dilaton field.
Here we introduce the unperturbed, state-dependent mass \mbox{$m_{j,0} \coloneqq m_j(0)$} and the dimensionless effective coupling parameter $\varepsilon_j \coloneqq \partial_\varrho \ln(m_j) \rvert_{\varrho=0}$.
In principle one can connect $\varepsilon_j$ to the individual constituents and natural constants of the atom, introducing coupling parameters that are independent of the atomic species.~\cite{Safronova2018}
While we refrain from such approaches and focus on signatures of DM in the detector signal, we emphasize that this discussion can be helpful for the design of the sensor and the choice of the atomic source.

\begin{figure}[htb!]
    \centering
    \includegraphics[scale=1]{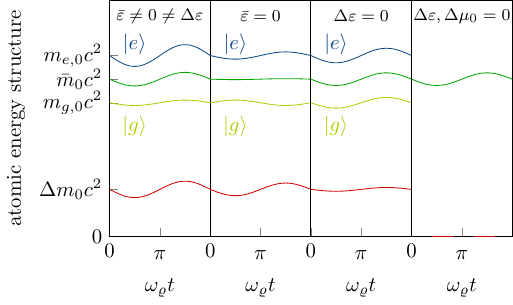}
    \caption{%
    The atomic energy structure oscillates around its Standard-Model values due to the oscillating part of the dilaton field, as apparent from Eq.~\eqref{eq:dilaton_field}. %
    The atomic-energy levels of the ground state (light green solid line), exited state (blue solid line), the atom's mean mass (green solid line), and energy difference between both states (red solid line), which corresponds to the mass defect, oscillate and are plotted against the phase $\omega_\varrho t$. %
    The relevant parameters are the mean coupling $\bar \varepsilon$ and differential coupling $\Delta \varepsilon$ of the internal states to the dilaton field, as well as the atom's mass defect $\DeltaM = \Delta m_0/\bar{m}_0$. %
    Due to the oscillation of the internal energies, the mean mass as well as the atomic transition frequency also oscillate. %
    For single-photon-like schemes the three relevant situations are shown from left to right: %
    (i) The general case in which the mean and differential coupling are non-vanishing, \ie{} $\bar \varepsilon \neq 0 \neq \Delta \varepsilon$. %
    (ii) The internal states couple exactly opposite to the dilaton field, \ie{} $\bar \varepsilon = 0$. %
    Only this case leads to a constant mean mass. %
    (iii) Both internal states can couple identically to the dilaton field, \ie{} $\Delta \varepsilon = 0$. %
    For Bragg-type schemes we have $\DeltaM = 0 = \Delta \varepsilon$ (right panel), as no internal transition is relevant. In this case we find only an oscillating mean mass that affects the motion of the atom.%
    }
    \label{fig:mass_oscil}
\end{figure}

The (dimensionless) dilaton field~\cite{DiPumpo2022, Hees2016}
\begin{equation}
\label{eq:dilaton_field}
    \varrho(z, t) = \varepsilon_S g_0 z/c^2 + \varrho_0 \cos \! \left( \omDM{} t - k_\varrho z + \phi_\varrho \right),
\end{equation}
includes the dimensionless coupling constant $\varepsilon_S$ of the source mass.
The dilaton field consists of two contributions:
(i) The first part introduces modifications of the gravitational potential leading to EP violations.~\cite{DiPumpo2023}
For example, it implies that the gravitational acceleration may be state-dependent, providing hints to extend general relativity.
(ii) The second part is an oscillating background field which can model cosmic DM.~\cite{Hees2018}
It behaves like a plane wave with the amplitude $\varrho_0$ and wave vector $k_\varrho$ at frequency $\omDM$.
The initial phase $\phi_\varrho$ is unknown, so that only a stochastic background of cosmic DM will be observed by the detector.
Through its coupling to the energies of the individual states and by that through mass-energy equivalence to the mass of the atom, see Fig.~\ref{fig:mass_oscil}, this oscillating background field directly influences both the COM motion and the atomic transition frequency.
In turn, such effects induce signatures of DM in the detector's signal.

Incorporating $k_\varrho \neq 0$ is not straightforward since one has to perform a non-relativistic limit of a dilaton-modified field theory to avoid operator-ordering issues.
However, galactic and cosmic observations~\cite{Tulin2018, Geraci2016, Arvanitaki2018} suggest small momenta compared to the rest mass of the dilaton, such that we assume $k_\varrho = 0$ in the following.
Thus, the dilaton field's frequency reduces to its Compton frequency $\omDM = m_\varrho c^2/\hbar$ and is solely determined by its mass $m_\varrho$.
In this case, the dimensionless dilaton amplitude~\cite{Hees2016, Filzinger2023} becomes
\begin{equation}
\label{eq:varrho_0}
    \varrho_0 = \frac{m_\mathrm{P} c^2}{\hbar \omDM} \sqrt{8 \pi \frac{\varrho_\mathrm{DM} L_\mathrm{P}^3}{m_\mathrm{P} c^2}},
\end{equation}
where the DM energy density $\varrho_\mathrm{DM} \cong \SI{0.4}{\giga \electronvolt \per \centi \meter^3}$ is compared to the energy scale given by the Planck mass $m_\mathrm{P} = (\hbar c/ G)^{1/2}$ and a volume defined by the Planck length $L_\mathrm{P} = (\hbar G/ c^3)^{1/2}$, with the gravitational constant $G$.

In addition, one could model the coupling of the dilaton field to the source of the gravitational field by yet another dilaton field, possibly incoherent to the one interacting with the atoms.
However, we assume that only one field is present, but allow for a phase shift $\phi_S$ compared to the oscillation of the dilaton field.
Hence, we model the time-dependent modulation of the gravitational acceleration via
\begin{equation}
    f(t) = \varrho_0 \cos(\vartheta + \phi_S),
\end{equation}
where $\vartheta \coloneqq \omDM t + \phi_\varrho$ is the phase of the dilaton field interacting with the atoms.
For example, a phase shift of $\pi/2$ corresponds to an oscillation of the atoms and gravitational acceleration out of phase.
We still can include incoherence between the dilaton-induced atomic properties and a gravitational modification by averaging independently over $\phi_S$ and $\phi_\varrho$ in analogy to the treatment discussed below.

\section{Dark-matter-induced Perturbations on Atoms}
With these insights, we expand the Hamiltonian from Eq.~\eqref{eq:start_hamil} with respect to $\varepsilon_j$, the unperturbed mass defect $\Delta m_0=m_{e,0}-m_{g,0}$, and $1/c^2$.
As a result, in first order the state-dependent mass can be replaced by
\begin{equation}
    m_j(t) \coloneqq \bar m_0 \left(1 + \MeanMDM(t) + \lambda_j\frac{ \DeltaM + \DeltaMDM(t) }{2} \right),
\end{equation}
including the unperturbed mean mass $\bar m_0=(m_{e,0} + m_{g,0})/2$ and different dimensionless modifications summarized in Table~\ref{tab:pert_parameters} with $\lambda_{e/g} = \pm 1$ to denote state-dependent perturbations.
We observe the following effects:
The mean mass oscillates due to $\MeanMDM(t)$ and thereby influencing the COM motion.
The dimensionless mass defect $\DeltaM$ gets further modified by $\DeltaMDM(t)$, which results in the oscillation of the transition energy explained below.

\begin{table}[h!]
\caption{\label{tab:pert_parameters}%
Summary of all parameters that give rise to the atomic structure and dilaton-induced perturbations. %
We define mean and differential values for the unperturbed mass and coupling parameters through the respective quantities of the individual states, as well as the dimensionless mass defect.
The dilaton field induces an oscillation with amplitude $\varrho_0$ and phase $\vartheta = \omDM t + \phi_\varrho$ of the mean mass, the energies of the individual states, as well as gravity, included in the dimensionless parameters $\MeanMDM(t)$, $\DeltaMDM(t)$, and $\MeanGDM(t)$, respectively.
For quantities already including one perturbative parameter, the contribution in braces must be neglected for consistency.
Moreover, we introduce the dilaton's coupling $\varepsilon_S$ to the gravitational source mass, which leads to mean and differential EP violations. %
}
\renewcommand{\arraystretch}{1.2}
\begin{ruledtabular}
\begin{tabular}{lcr}
Cause & Parameter & Definition\\
\hline
mean mass & $\bar m_0$ & $(m_{e,0} + m_{g,0})/2$ \\
mass defect & $\Delta m_0$ & $m_{e,0} - m_{g,0}$ \\
mass defect &$\DeltaM$ & $\Delta m_0/ \bar m_0$ \\
mean coupling & $\bar \varepsilon$ & $(\varepsilon_{e} + \varepsilon_{g})/2$ \\
differential coupling & $\Delta \varepsilon$ & $\varepsilon_{e} - \varepsilon_{g}$ \\
mean-mass osci.&$\MeanMDM(t)$ & $\varrho_0 [\bar{\varepsilon} + \{\Delta m_0 \Delta \varepsilon /(4\bar{m}_0)\}] \cos\vartheta$ \\
state-dep. osci.&$\DeltaMDM(t)$ & $\varrho_0 [\Delta \varepsilon + \{\Delta m_0 \bar{\varepsilon}/\bar{m}_0\}] \cos\vartheta$ \\
osci. of gravity&$\MeanGDM(t)$ & $\varepsilon_S \varrho_0 \cos(\vartheta + \phi_S)$ \\
mean-mass EP viol. &$\MeanGEP$ & $\varepsilon_S \bar{\varepsilon}$ \\
state-dep. EP viol. &$\DeltaGEP$ & $\varepsilon_S \Delta \varepsilon$ \\
\end{tabular}
\end{ruledtabular}
\end{table}

Furthermore, the dilaton field effectively leads to a modification of the gravitational acceleration $g_0$ which becomes state dependent, \ie{}
\begin{equation}
    g_j(t) \coloneqq g_0 \left(1 +\MeanGEP + \MeanGDM(t) + \lambda_j \frac{\DeltaGEP}{2}\right),
\end{equation}
where $\MeanGEP$ parametrizes EP violations between different atomic species depending on the mean coupling of the dilaton field to the atom.~\cite{Schlippert2014, DiPumpo2021, Barrett2022}
An EP violation between different internal states~\cite{Zhang2020, DiPumpo2023} is encoded in $\DeltaGEP$.
In addition, the gravitational acceleration changes dynamically via $\MeanGDM(t)$ through the DM coupling to the source mass.

We separate the resulting Hamiltonian 
\begin{equation}
\label{eq:H(t)}
    \hat H(t)  = \sum_{j=g,e} \left[ \hat{H}_0 + \hat{V}_m(t) + \hat{V}_\mathrm{kin}(t) + \hat{V}_\mathrm{pot}(t) \right] \otimes \ket{j} \bra{j},
\end{equation}
into an unperturbed part $\hat{H}_0$ and perturbations $\hat{V}_m$ of the rest mass, $\hat{V}_\mathrm{kin}$ of the kinetic energy, and $\hat{V}_\mathrm{pot}$ of  the potential energy.
Besides, we generalize $\lambda_j$ to $\lambda_j(t)$ to allow for time-dependent changes of the internal state.
We discuss the explicit form of these four contributions in the following:

(i) The unperturbed Hamiltonian
\begin{equation}
\label{eq:unpert_hamil}
    \hat{H}_0 = \bar{m}_0 c^2 + \frac{\hat{p}^2}{2 \bar{m}_0} + \bar{m}_0 g_0 \hat{z},
\end{equation}
describes a particle of mean mass $\bar m_0$ moving in a linear gravitational potential without any state-dependent effects or internal structure.

(ii) The rest-mass perturbation
\begin{equation}
\label{eq:V_m}
    \hat{V}_m \coloneqq \bar m_0 c^2 \left[\MeanMDM(t) + \lambda_j(t)\frac{\DeltaM + \DeltaMDM(t)}{2} \right],
\end{equation}
does not affect the motion of the atom, but changes the Compton frequency $\omega_c \coloneqq \bar m_0 c^2/ \hbar$ and atomic transition frequency $\Omega \coloneqq \Delta m_0 c^2/ \hbar$.
The phase measured by atomic clocks, Mach-Zehnder and comparable interferometers is not affected by modifications of the Compton frequency to lowest order.%
\footnote{
    The Compton frequency is modified to $\omega_c \mapsto \bar m_0 c^2/\hbar[ 1 + \MeanMDM(t)] = \omega_c + ( \omega_c \bar \varepsilon +  \Omega \Delta \varepsilon/4 ) \varrho_0 \cos \vartheta$.
    If both internal states couple identically to the dilaton field, \ie{}, $\Delta \varepsilon=0$, the change of the Compton frequency is directly proportional to $\omega_c$, in contrast to the atomic transition frequency.%
}
However, they are sensitive to the atomic transition frequency between both internal states, which is modified to
\begin{equation}
\label{eq:change_of_transfreq}
   \Omega \mapsto \frac{\bar m_0 c^2}{\hbar}[ \DeltaM + \DeltaMDM(t)] = \Omega + \delta \Omega \cos\vartheta.
\end{equation}
The unperturbed atomic transition frequency $\Omega$ is connected to the mass defect and is modulated by an oscillation with amplitude $\delta \Omega  \coloneqq ( \omega_c \Delta \varepsilon +  \Omega \bar \varepsilon ) \varrho_0$.
Here, $\bar \varepsilon$ is the mean coupling of both internal states to the dilaton, whereas $\Delta \varepsilon$ is the difference of their coupling constants, see Table~\ref{tab:pert_parameters}.
Commonly, it is assumed that $\delta \Omega$ is proportional to the atomic transition frequency,~\cite{Arvanitaki2015, Arvanitaki2018, Badurina2022, Hees2018, Safronova2018} \ie{} the coupling of the dilaton field to both internal states is identical resulting in $\Delta \varepsilon=0$.
Since the details of the coupling are \emph{a priori} unknown, it is important to allow a different coupling of both internal states, \ie{} to allow for $\Delta \varepsilon \neq 0$.
Generally, $\delta \Omega$ is a linear combination of both the Compton and the atomic transition frequency with respect to the different coupling parameters.
While the coupling parameters $\Delta \varepsilon$ and $\bar \varepsilon$ may be of very different orders, the Compton and atomic transition frequency also differ by multiple orders of magnitude so that in principle both contributions to $\delta \Omega$ have to be considered.
Remarkably, the change of the atomic transition frequency depends on the Compton frequency, which could enhance the DM signature in the detector.
Besides, an observable change in the atomic transition frequency present for Raman~\cite{Zhan2020} or microwave transitions cannot be completely ruled out.
Such transitions are relatively simple to implement and benefit from a much lower recoil velocity, suppressing the impact of gravity-gradient noise.

(iii) The kinetic-energy perturbation
\begin{equation}
\label{eq:V_kin}
    \hat{V}_\mathrm{kin} \coloneqq -\frac{\hat{p}^2}{2 \bar{m}_0} \left[\MeanMDM(t) + \lambda_j(t)\frac{ \DeltaM + \DeltaMDM(t) }{2} \right]
\end{equation}
is caused by changes in $m_j$ affects the COM motion.
The DM-induced oscillation of the mean mass leads to a time-dependent kinetic energy.
Additional state-dependent contributions arise from the mass defect \mbox{$\DeltaM > 0$}, \ie{} the ground state gains kinetic energy while the excited state loses kinetic energy.
Further state-dependent mass oscillations arise due to the dilaton field encoded in $\DeltaMDM(t)$.

(iv) The potential-energy perturbation 
\begin{equation}
\label{eq:V_pot}
\begin{split}
    \hat{V}_\mathrm{pot} \coloneqq \bar{m}_0 g_0 \hat{z}\bigg[ &\MeanGEP + \MeanGDM(t) + \MeanMDM(t) \\
    &+ \lambda_j(t)\frac{\DeltaGEP + \DeltaM + \DeltaMDM(t)}{2}  \bigg]
\end{split}
\end{equation}
due to changes in both $m_j$ and $g_j$ also affects the COM motion, similar to $\hat{V}_\mathrm{kin}$.
However, we notice more contributions:
An additional shift $\MeanGEP$ of the gravitational acceleration due to the mean-mass EP violation occurs.
This shift is relevant for tests of the universality of free fall between different atomic species.
Similarly, differential accelerations between the atom in different internal states are encoded in $\DeltaGEP$, which is a different facet of possible EP violations, \eg{} relevant for tests of the universality of clock rates.
Further, we observe a global sign difference in $\hat{V}_\mathrm{kin}$ and $\hat{V}_\mathrm{pot}$ compared to the unperturbed case.
Thus, it raises the expectation that some perturbations contribute twice.

As a side note we mention that both $\hat{V}_\mathrm{kin}$ and $\hat{V}_\mathrm{pot}$ depend on the perturbation parameters $\MeanMDM(t)$ and $\DeltaMDM(t)$.
In principle, they include products of a dilaton coupling with $ \Delta m_0/ \bar m_0$, which are next order in perturbation.
Therefore, they are omitted in further calculations and enclosed in Table~\ref{tab:pert_parameters} by braces.

The discussed perturbations reflect themselves in the signals of atom interferometers that can be used to construct DM detectors.
In Secs.~\ref{sec:dm_signal_gradiometers} and \ref{sec:sp_interferometers}, we discuss the individual contributions and their order of magnitude in an exemplary setup.

\section{Dark-matter Signal in Atom Gradiometers}
\label{sec:dm_signal_gradiometers}
\begin{figure}[htb!]
    \centering
    \includegraphics[scale=1]{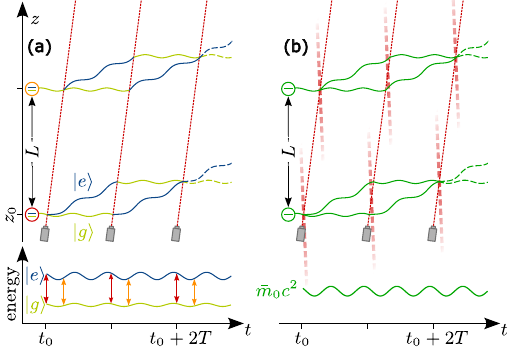}
    \caption{%
    Space-time diagram of dark-matter detectors based on Mach-Zehnder gradiometers implemented via {\sffamily\bfseries (a)} single-photon transitions and {\sffamily\bfseries (b)} two-photon transitions (Bragg). %
    The individual trajectories are shown in a freely falling frame for simplicity and are therefore solely influenced by the coupling of DM to the COM motion, which causes an (exaggerated) oscillation.
    Additionally, the lower panels show the oscillating energy scales, which also affects the rest-mass perturbation. %
    Initially, two wave packets in their ground state $\ket{g}$ are separated by the distance $L > 0$. %
    The first beam-splitter pulse splits the wave packet located at $(t_0, z_0)$ into two arms: %
    {\sffamily\bfseries (a)} An upper arm in the excited state $\ket{e}$ with increased momentum $\hbar k$ trough single-photon absorption and a lower arm in which the atoms are still in the ground state. %
    {\sffamily\bfseries (b)} Bragg-diffraction leaves the internal state unaffected but generates a superposition of two momenta through a two-photon process. %
    After an additional time $T$ a mirror pulse interchanges the roles of both arms. %
    Then, at the time $t_0 + 2T$ a second beam-splitter pulse interferes them. %
    The wave packet located at $z_0 + L$ undergoes the same process, but with a delay of $\tau_L = L/c$ due to the propagation time of light. %
    While the oscillating energy scales influence the motion of the atom, for single-photon absorption the atomic transition frequency is additionally probed at different times, visualized through arrows in the term diagram by arrows. %
    Due to the propagation delay, the transitions of the upper interferometer (orange) are shifted compared to the lower one (red).%
    }
    \label{fig:mzi_setup}
\end{figure}

Quantum sensors such as atomic clocks or atom interferometers are affected~\cite{Safronova2018} by the previously derived perturbations.
We focus on the signal observed by DM detectors generated from light-pulse atom interferometers.
Such high-precision quantum sensors have been proposed for DM searches,~\cite{Derevianko2014, Arvanitaki2015, Geraci2016, Arvanitaki2018, Badurina2023} whose signal can be enhanced by multi-diamond~\cite{Graham2016,Schubert2019, Schubert2021,dipumpo2023optimal} along with large-momentum-transfer techniques.~\cite{Chiow2011,Graham2013, Gebbe2021}
To focus on the fundamental effects on these detectors, we study an atomic Mach-Zehnder interferometer (MZI) \cf{} Fig.~\ref{fig:mzi_setup} without transferring large momenta.
However, the treatment introduced below can easily be generalized to different interferometer types and geometries.

In an MZI, the wave packet is initially split by a beam-splitter pulse into two separate arms.
One arm continues along the initial path, while the atoms on the other arm gain momentum due to diffraction, so that the arms become spatially separated.
After an interrogation time $T$, a mirror pulse interchanges the momenta of the two arms.
At the final time of $2T$, a second beam-splitter pulse interferes both arms, resulting in two output ports.
Each interaction of the atoms with light that transfers momentum might also change their internal state.
Here, the specific implementation is the key for the sensitivity of a detector.

In delta-pulse approximation~\cite{Schleich2013, Funai2019} the effect of light-matter interaction on the atomic motion is described by an effective arm-dependent potential~\cite{Ufrecht2020}
\begin{equation}
    \hat{V}_\mathrm{em} = - \hbar \sum_\ell k_\ell^{(j)} \hat{z} \delta(t-t_\ell),
\end{equation}
with $k_\ell^{(j)}$ being the $\ell^\mathrm{th}$ (effective) wave vector acting at time $t_\ell$, transferring the momentum $\hbar k_\ell^{(j)}$ on arm $j$, and $\delta(\cdot)$ denoting the Delta distribution.
We assume that both arms are diffracted at the same time, neglecting the finite speed of light on the scale of the arm separation.~\cite{Tan2017_Time}
Moreover, we do not include frequency chirps necessary in terrestrial, vertical configurations and omit the standard laser-phase contribution. 

This interaction allows for different momentum-transfer mechanisms and can in principle incorporate large momentum transfer.
Typically, the interaction is categorized into single-photon~\cite{Bott2023} and (counter-propagating) two-photon~\cite{Hartmann2020} transitions.
In the former, $k_\ell^{(j)}$ simply corresponds to the laser's wave vector, while in the latter, $k_\ell^{(j)}$ is the effective wave vector given by the sum of both lasers' wave vectors.
Two-photon transitions suffer from laser-phase noise in differential setups with long baselines, which is suppressed for single-photon transitions.
Nevertheless, two-photon transitions are generally more flexible.
They allow to only transfer momentum without changing the internal state (Bragg) and drive effectively hyperfine-structure transitions while transferring momentum that corresponds to optical wavelengths (Raman).

Assuming $ |k_\ell^{(j)}|=k$ for all interaction times and vanishing gravity gradients leads to a closed, unperturbed MZI.
Deviations are introduced by the dilaton field as discussed in Eqs.~\eqref{eq:V_m}--\eqref{eq:V_pot}.
In a perturbative treatment,~\cite{Ufrecht2020} we find to first order the phase
\begin{equation}
    \phi (z_0, p_0, t_0) = -k g_0 T^2 + \phi_m + \sum_{i=1}^{14}  \phi_i,
\end{equation}
with the wave packet's initial position $z_0$, as well as initial momentum $p_0$, %
depending on the time $t_0$ at which the interferometer is initialized.
It includes the standard phase $-k g_0 T^2$ of an MZI, which is perturbed by the contributions $\phi_m$ and $\phi_i$ induced by the dilaton field, %
compare Table~\ref{tab:mzi_phases} in Appendix~\ref{sec:appendix}  for their explicit expressions.

Common-mode operation of two MZIs spatially separated by the distance $L > 0$, \eg{} where the first MZI is located at $z_0$ and the second one at $z_0 + L$, suppresses noise and the dominant phase $-k g_0 T^2$ for vanishing gravity gradients.~\cite{Graham2013}
We account for the finite speed of light on the separation scale of the two interferometers, but not on the extent of the arm separation of a single one.
The first MZI starts its sequence at time $t_0$ with an initial momentum $p_0$ of the wave packet, while the second interferometer starts at time $t_0 + L/c$ with an initial momentum $p_1$.
Subtracting the phases of both interferometers gives rise to the differential phase, which in our approximation only depends on dilaton-introduced perturbations
\begin{equation}
    \Delta \phi = \Delta \phi_m + \sum_{i=1}^{14} \Delta \phi_i,
\end{equation}
with $\Delta \phi_i \coloneqq \phi_i(z_0 + L, p_1, t_0 + \tau_L) - \phi_i(z_0, p_0, t_0)$ and the propagation delay $\tau_L \coloneqq L/c$ of the light pulse.

Since the dilaton phase $\phi_\varrho$ may vary, we can only measure this stochastic background and we therefore have to average.
Thus, we assume a uniform distribution of $\phi_\varrho$ in accordance with the principle of maximum entropy.~\cite{Jaynes1957}
The signal amplitude~\cite{Arvanitaki2018}
\begin{equation}
\label{eq:signal_contrib}
    \Phi_{\mathrm{S}}^2 = 2 \int\limits_{0}^{2\pi} \frac{\! \dd \phi_\varrho}{2 \pi} \sum_{i,j} \Delta \phi_i(\phi_\varrho) \Delta \phi_j(\phi_\varrho) %
    \eqqcolon 2 \sum_{i,j} \big\langle \Delta \phi^2_{i,j}\big\rangle
\end{equation}
is the square of the phase difference $\Delta \phi$ averaged over $\phi_\varrho$.
Due to the square, various correlations $\big\langle  \Delta \phi^2_{i,j} \big\rangle$ of the individual phases contribute.

In the following we identify dominant contributions to the signal, \ie{} $\big\langle \Delta \phi^2_{i,j} \big\rangle$, for different experimental realizations, namely for single-photon-type and Bragg-type interferometers.

\section{Single-Photon Interferometers}
\label{sec:sp_interferometers}
Many planned DM detectors~\cite{Abe2021, Badurina2020} based on atom interferometry rely on single-photon transitions, due to their intrinsic suppression of laser-phase noise.~\cite{Yu2011, Graham2016} 
In this section, we consider the effects of the dilaton field on such interferometers and focus on the dominant contributions of the observed signal amplitude.
Using single-photon transitions for atomic diffraction not only changes the momentum of the atom, but also its internal state.
The results discussed in this section therefore also transfer to Raman transitions, also planned for some sensors,~\cite{Zhan2020} where only the frequency scales have to be adjusted.

We first focus on the phase difference introduced by the modified atomic transition frequency \cf{} Eq.~\eqref{eq:change_of_transfreq} which gives rise to
\begin{equation}
    \Delta \phi_m = - \varrho_0 (\Delta \varepsilon \omega_c + \bar{\varepsilon} \Omega) \left[ \tau_1(t_0 + \tau_L) - \tau_1(t_0) \right],
\end{equation}
where the timescale $\tau_1(t_0)$ is proportional to $1/\omDM$ and is listed in Table~\ref{tab:time_scales} in Appendix~\ref{sec:appendix}.
Recalling $\varrho_0 \propto 1/\omDM$ from Eq.~\eqref{eq:varrho_0} shows that this contribution plays an important role in the search for ultralight DM.
In particular, the change of the atomic transition frequency yields the dominant contribution
\begin{equation}
\label{eq:DeltaPhi_mm}
    \big\langle \Delta \phi^2_{m,m} \big\rangle = 32 \frac{(\Delta \varepsilon \omega_c + \bar{\varepsilon} \Omega)^2}{\omDM^2} \varrho_0^2 \SigSin{\tau_L}{2} \SigSin{T}{4},
\end{equation}
to the signal amplitude, with $\SigSin{t}{1} \coloneqq \sin \!\left(\omDM t/2\right)$.
We refer to factors like $\mathcal{S}$ as interferometric factors that include the interrogation-mode function.~\cite{dipumpo2023optimal}
Perturbations and scaling factors like $\varrho_0^2 (\Delta \varepsilon \omega_c + \bar{\varepsilon} \Omega)^2/\omDM^2$ are independent of the explicit interferometer geometry and appear in a similar form for different geometries, including those with large momentum transfer.

Equation~\eqref{eq:DeltaPhi_mm} is a generalization of other treatments of atom-interferometric DM detectors~\cite{Arvanitaki2018} but has a similar form.
In fact, the differential coupling $\Delta \varepsilon$ is usually neglected, which introduces another frequency scale.
In principle, this contribution also arises for Raman-type interferometers, where the atomic energy difference is in the microwave range and much smaller than for optical transitions.
However, due to the coupling to the Compton frequency that is orders of magnitude larger, one can also expect a sensitivity to the parameter $\Delta \varepsilon$ for Raman setups.
The same holds true for single-photon transitions between hyperfine states in the microwave range.
For this type of transitions, in contrary, the momentum transfer is negligible, resulting in interferometers that are less sensitive to gravity-gradient noise.

As discussed above, generally the coupling scheme is unknown so that both $\Delta \varepsilon$ and $\bar \varepsilon$ might contribute.
Since their order of magnitude is unknown,  we consider in the following two limiting cases:
Either the coupling of both internal states is completely identical, \ie{} $\Delta \varepsilon = 0$, or exactly opposite, \ie{} $\bar \varepsilon = 0$.

\subsection{Vanishing Mean Coupling ($\bar \varepsilon = 0$)}
\label{subsec:vanishing_mean_coupling}
For vanishing mean coupling, the change of the atomic transition frequency reduces to $\delta \Omega = \omega_c \Delta \varepsilon \varrho_0$.
Hence, the relevant scale is given by the Compton frequency and \emph{not} by the atomic transition frequency. 
This is different to what is usually postulated in most treatments for AIs and clocks, where $\delta \Omega \propto \Omega$ is assumed.
Since $\omega_c \gg \Omega$ generally applies, we expect a larger suppression of contributions which do not originate from the clock phase but keep in mind that we probe for a different coupling parameter.
For example, strontium is a promising candidate~\cite{Loriani2019} for future single-photon AIs~\cite{Hu2017, Rudolph2020} and gives rise to $\Omega / \omega_c \cong \SI{e-11}{}$.

Consequently, we neglect all $\big\langle \Delta \phi^2_{i,j} \big\rangle$ for $i,j \neq m$ as their scale factors have to be compared to the Compton frequency, and arrive at the signal amplitude
\begin{equation}
\label{eq:signal_bareps_0}
    \Phi_\mathrm{S}^2 \cong 64 \frac{\omega_c^2}{\omDM^2} \Delta \varepsilon^2 \varrho_0^2 \SigSin{\tau_L}{2} \SigSin{T}{4}.
\end{equation}
We recognize the twofold effect of the Compton frequency:
On the one hand it leads to the suppression of phase contributions competing with $\Delta \phi_m$.
And on the other hand, it enhances the signal.
This result also persists for setups where small transition frequencies are used, \eg{} hyperfine or Raman transitions.
In this case, the Compton frequency clearly sets the relevant scale, even though the coupling parameter $\Delta \varepsilon$ might be small.

\vspace{-2mm}
\subsection{Vanishing Differential Coupling ($\Delta \varepsilon=0$)}
\label{subsec:vanishing_diff_coupling}
If both internal states couple identically to the dilaton field, as assumed in most previous treatments, the modulation amplitude of the atomic transition frequency takes the form $\delta \Omega = \Omega \bar \varepsilon \varrho_0$.
Hence, the relevant scale is now the atomic transition frequency.
For single-photon transitions, the atomic transition frequency benefits from an optical regime and has a clear advantage over hyperfine or Raman transitions.
Similar to the previous discussion,
\begin{equation}
\label{eq:phi_mm_deltaeps_0}
    2\big\langle \Delta \phi^2_{m,m} \big\rangle = 64 \frac{\Omega^2}{\omDM^2} \bar \varepsilon^2 \varrho_0^2 \SigSin{\tau_L}{2} \SigSin{T}{4}
\end{equation}
is the dominant contribution.
But, by decreasing the relevant scale by several orders of magnitude, we include $\big\langle \Delta \phi^2_{m,j} \big\rangle$ as next-order contributions to the signal amplitude.

After averaging over $\phi_\varrho$, the surviving next-order contributions to the signal amplitude $\big\langle \Delta \phi^2_{m,1} \big\rangle$, $\big\langle \Delta \phi^2_{m,2} \big\rangle$, and $\big\langle \Delta \phi^2_{m,9} \big\rangle$ (listed in this order) give rise to
\begin{equation}
    \frac{\Phi_\mathrm{S}^2}{2 \big\langle \Delta \phi_{m,m}^2 \big\rangle} \cong 1 %
    - \frac{\omega_k}{\Omega} (1+4\bar\wp_0)
    + 2 \frac{k g_0 T}{\Omega}%
    - 2 \frac{k g_0}{\Omega \omDM} \frac{\varepsilon_S}{\bar \varepsilon} \sin \phi_S,
\end{equation}
with the recoil frequency $\omega_k = \hbar k^2/(2 \bar m_0)$ and the initial mean momentum $\bar \wp_0 \coloneqq (p_1 + p_0)/(2 \hbar k)$.

The dominant part of the signal, \ie{} $\big\langle \Delta \phi^2_{m,m} \big\rangle$, has already been discussed.~\cite{Arvanitaki2018}
We provide the next sub-leading corrections and observe that even in spaceborne experiments $\big\langle \Delta \phi^2_{m,1} \big\rangle$ provides a purely kinetic contribution.
Additionally, for terrestrial setups long interrogation times $2T$ increase the significance of $\big\langle \Delta \phi^2_{m,2} \big\rangle$.
The contribution $\big\langle \Delta \phi^2_{m,9} \big\rangle$ induced by oscillating gravity vanishes for an in-phase oscillation, \ie{} $\phi_S = 0$, but can be enhanced by $\phi_S = \pm \pi/2$.

\vspace{-2mm}
\subsection{Influence of Both Couplings}
\begin{figure}[t!]
    \centering
    \includegraphics[scale=1]{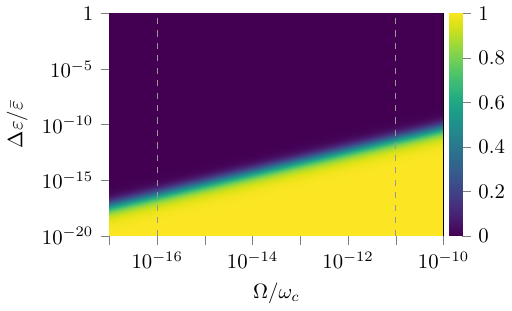}
    \caption{%
    Dominant contribution to the signal amplitude for the case considered in most treatments, \ie{} with $\Delta \varepsilon=0$, compared to the general expression that also allows for a differential coupling.
    The figure shows the fraction  $\big\langle \Delta \phi^2_{m,m} \big\rangle \big\vert_{\Delta \varepsilon=0}\big/ \big\langle \Delta \phi^2_{m,m} \big\rangle$ with $\Delta \varepsilon / \bar \varepsilon > 0$. %
    For reference, we have highlighted the frequency ratios for single-photon transitions ($\Omega/\omega_c \cong \SI{e-11}{}$ for clock transitions in strontium) and for Raman-type schemes  ($\Omega/\omega_c \cong \SI{e-16}{}$ for hyperfine transitions in rubidium) by vertical lines. %
    The bottom right corresponds to the standard case described in Sec.~\ref{subsec:vanishing_diff_coupling}, while the upper left corresponds to the regime introduced in Sec.~\ref{subsec:vanishing_mean_coupling}.
    The figure highlights the transition between both regimes, where the general expression has to be taken into account.
    Even if the ratio $\Delta \varepsilon / \bar \varepsilon$ is small, this difference can be compensated due to the different order of magnitude of the transition frequency and the Compton frequency.%
    }
    \label{fig:no_diff_coupling_comparison}
\end{figure}

Since one usually neglects the differential coupling $\Delta \varepsilon$, we briefly discuss its influence for different values of $\Omega /\omega_c$.
For that, we compare this approximation to the full expression by studying the ratio $\big\langle \Delta \phi^2_{m,m} \big\rangle \big\vert_{\Delta \varepsilon=0} \big/ \big\langle \Delta \phi^2_{m,m} \big\rangle$.
It is plotted in  Fig.~\ref{fig:no_diff_coupling_comparison} as a function of  $\Omega /\omega_c$ and $\Delta \varepsilon / \bar \varepsilon >0$.
It shows a drastic change between the two regimes discussed above.
However, Fig.~\ref{fig:no_diff_coupling_comparison} gives a hint where the general signal-amplitude contribution has to be considered in the analysis for an expected range of coupling parameters and depending on the specific atomic species and transition frequency.
The figure highlights that even if the ratio $\Delta \varepsilon / \bar \varepsilon$ is small, it can be compensated due to the different order of magnitude of the transition frequency and the Compton frequency.

\section{Bragg-Type Interferometers ($\DeltaM = 0 = \Delta \varepsilon$)}
Finally, we turn to Bragg-type MZIs, where a two-photon process is used to only transfer momentum without changing the internal state.
Therefore, we have $\DeltaM=0=\Delta \varepsilon$, which directly implies that such interferometers are only susceptible to the mean coupling $\bar \varepsilon$ of the atom to the dilaton field.
While there have been proposals for DM detectors that focus on the COM motion~\cite{Geraci2016} and not on the internal structure of the atom, which effectively corresponds to Bragg-based setups, differential configurations have not yet been discussed in detail.
Currently, in the context of very-long-baseline atom interferometry differential setups, which can in principle be used for DM detection, are envisioned to rely on Bragg diffraction, \eg{} MAGIS-100,~\cite{Abe2021} MIGA,~\cite{Canuel2018} or ZAIGA.~\cite{Zhan2020} 

For Bragg-type interferometers only a few phase contributions remain in the differential setup \cf{} Table~\ref{tab:mzi_phases} in Appendix~\ref{sec:appendix}, some of which have been also calculated before.~\cite{Geraci2016}
These terms also arise for single-photon transitions, where, as previously discussed, the clock contribution dominates.
Among these remaining phase contributions, it is not straightforward to determine which one is the dominant one in the signal amplitude.
We therefore list all of them in Table~\ref{tab:bragg_signal_contributions} in Appendix~\ref{sec:appendix}.
With the help of this overview one can identify three different scales in the signal amplitude, independent of the violation parameters and interferometric factors.
For each experiment it has to be checked individually which scale is the dominant one.
However, the limiting case $g_0 = 0$, which is important for spaceborne experiments or horizontal setups like the MIGA project, simplifies the signal amplitude to
\begin{equation}
\label{eq:signal_bragg_nograv}
    \Phi_\mathrm{S}^2 = 32 \frac{\omega_k^2}{\omDM^2} {\bar\varepsilon}^2 \varrho_0^2 \times \text{intf. factor},
\end{equation}
where the interferometric factor originates from $\big\langle \Delta \phi^2_{1,1} \big\rangle$, which is the only remaining contribution \cf{} Table~\ref{tab:bragg_signal_contributions}.
Since all other terms vanish, these types of setups are less susceptible for ultralight DM searches compared to terrestrial setups, where gravity-induced contributions are present.

In summary, we can identify different relevant scales in the various limiting cases by comparing Eqs.~\eqref{eq:signal_bareps_0}--\eqref{eq:signal_bragg_nograv}.
This comparison leads to $\omega_c \gg \Omega \gg \omega_k$.
We see that the single-photon-type interferometers benefit from the Compton and atomic transition frequency, whereas Bragg-type interferometers in zero gravity only depend on the recoil frequency.

\section{Conclusions}
In our article, we demonstrated that assuming a linear scaling of the dark-matter-induced change of the atomic transition frequency with the unperturbed transition frequency is only an approximation for vanishing differential coupling that is made in most discussions.~\cite{Geraci2016, Arvanitaki2018, Badurina2023}
Atom interferometers that rely on the change of internal states, \eg{} single-photon or Raman transitions, are susceptible to both types of coupling, where the signal is dominated by clock-type phases.~\cite{Norcia2017}
In these cases, the atomic transition and Compton frequency are relevant frequency scales weighted by the mean and differential coupling of both states to dark matter, respectively.
However, Bragg-type atom interferometers that preserve the internal state are less susceptible to dark matter in the sense, that they only depend on the mean coupling and the recoil frequency.
Therefore, they can only measure effects on the motion like accelerometers.~\cite{Graham2016,Geraci2016}

Our results support the dark-matter search with single-photon atom interferometers in terrestrial setups since their signal profits from any dark-matter-related contributions.
Nevertheless, a detailed noise analysis is necessary.
Additionally, various mechanisms for atomic diffraction can be used to isolate different coupling parameters, \eg{} the results from spaceborne Bragg-type setups can give bounds on the mean coupling.
These limits can be combined with the results from other setups to give bounds to both independent coupling parameters and make connections to the tests of the Einstein equivalence principle.

In perspective, a generalization of our treatment to different geometries might be useful to boost dark-matter signatures in the signal, \eg{} large-momentum-transfer techniques.
In this context the effect of the modified Compton frequency can be studied in recoil measurements for dark-matter detection or Ramsey-Bord\'e-type interferometers.~\cite{Borde1989}
Following current developments in gravitational-wave detection, various differential schemes can be considered and atom-interferometer-based networks~\cite{Badurina2020, Figueroa2021} can be simultaneously used for dark-matter searches and the detection of gravitational waves.
Finally, implications for quantum-clock interferometry~\cite{Loriani2019, Zych2011} can be studied, \ie{} propagating a superposition of internal states along each interferometer arm, to focus on the effect of the different degrees of freedom.


\begin{acknowledgments}
We thank F. Di Pumpo, A. Friedrich, A. Geyer, C. Ufrecht, as well as the QUANTUS and INTENTAS teams for fruitful and interesting discussions.
The QUANTUS and INTENTAS projects are supported by the German Space Agency at the German Aerospace Center (Deutsche Raumfahrtagentur im Deutschen Zentrum f\"ur Luft- und Raumfahrt, DLR) with funds provided by the Federal Ministry for Economic Affairs and Climate Action (Bundesministerium f\"ur Wirtschaft und Klimaschutz, BMWK) due to an enactment of the German Bundestag under Grant Nos. 50WM2250E (QUANTUS+), as well as 50WM2177 (INTENTAS).
E.G. thanks the German Research Foundation (Deutsche Forschungsgemeinschaft, DFG) for a Mercator Fellowship within CRC 1227 (DQ-mat).
\end{acknowledgments}

\section*{Author Declarations}
\subsection*{Conflict of Interest}
The authors have no conflicts to disclose.

\subsection*{Author Contributions}
{\sffamily\textbf{Daniel Derr}} Conceptualization (supporting); Formal analysis (lead); Investigation (lead); Validation (equal); Visualization (lead); Writing - original draft (lead); Writing - review and editing (equal).

{\sffamily\textbf{Enno Giese}} Conceptualization (lead); Formal analysis (supporting); Funding acquisition (lead); Investigation (supporting); Supervision (lead); Validation (equal); Visualization (supporting); Writing - original draft (supporting); Writing - review and editing (equal).

\section*{Data Availability}
The data that support the findings of this study are available within the article.

\appendix
\section{Definitions, Phases and Signal Contributions}
\label{sec:appendix}
In this appendix, we summarize the main results and various contributions to the interferometer phase, as well as the signal amplitude.
To keep the description as simple as possible, we have collected these expressions in tables that are given here, to keep the main body of our article focused on the main results.
Nevertheless, the exact equations are provided here for completeness and can be used for further studies.

The phase contributions $\phi_j$ induced by dark matter into the interference signal of Mach-Zehnder interferometers are listed in Table~\ref{tab:mzi_phases}.
Here, we connect the individual phase contributions to the respective perturbative potential and identify the individual cause.
For a compact notation of these expressions, we introduce different time scales $\tau_i$ defined in Table~\ref{tab:time_scales}.

These phases serve for the calculation of the signal amplitude, where the variance of the phase differences between two interferometers in a gravimeter is averaged over the phase of the dilaton field.
Table~\ref{tab:sp_signal_contributions} summarizes all non-vanishing sub-leading contributions $\big\langle \Delta \phi^2_{m,j} \big\rangle$ to the signal amplitude of single-photon-like Mach-Zehnder interferometers.
Besides a numerical factor, we identify the respective scale, violation parameters, as well as an interferometric factor that depends on the geometry used, the initial momenta and the interrogation-mode function~\cite{dipumpo2023optimal}.
The same analysis is performed in Table~\ref{tab:bragg_signal_contributions} for the case of Bragg-type Mach-Zehnder interferometers.
Here, all possible contributions to the signal amplitude are given, as no dominant energy and frequency scale can be identified.

\begin{table*}
\caption{%
\label{tab:mzi_phases}%
Dilaton-field-induced perturbations in a two-level atom and their respective phase contributions $\phi_i$ in a Mach-Zehnder interferometer. %
Perturbations to the rest mass ($\hat{V}_m$), to the kinetic energy ($\hat{V}_\mathrm{kin}$), and to the potential energy ($\hat{V}_\mathrm{pot}$) are divided into different expressions due to their physical origin: the mass defect $\DeltaM$, the mean-mass oscillation $\MeanMDM(t)$, the state-dependent-mass oscillation $\DeltaMDM(t)$, the mean-mass EP violation $\MeanGEP$, the state-dependent EP violation $\DeltaGEP(t)$, and the oscillation of gravity $\MeanGDM(t)$. Their explicit definitions are summarized Table~\ref{tab:pert_parameters}. %
Each of these perturbations introduces a contribution $\phi_j$ to the interferometer phase, that is calculated perturbatively.~\cite{Ufrecht2020} %
Here, we introduced dimensionless momenta %
$\wp_0 \coloneqq p_0/(\hbar k)$, %
and $\bar \wp_{t_0+T} \coloneqq (p_0 - \bar m_0 g_0 T + \hbar k/2)/(\hbar k)$, which depend on the initial momentum $p_0$ the initialization time $t_0$ of the interferometer, as well as the (effective) wave vector $k$. %
For compact expressions, we use the time scales $\tau_1(t_0)$, $\tau_2(t_0)$, $\tau_3(t_0)$, $\tau_S(t_0)$, and $\tau_\mathrm{EP}(t_0)$ introduced in Table~\ref{tab:time_scales}.%
}
\renewcommand{\arraystretch}{2}
\begin{ruledtabular}
\begin{tabular}{lccrcl}
    Perturbative Potential & Origin & Expression & Abbreviation & & Phase Contribution\\ \hline
    \multirow{3}{*}{$\hat{V}_m$} & mean-mass oscillation & $\MeanMDM(t) \bar m_0 c^2$ & & & $\phantom{-} 0$ \\ \cline{2-6}
    & mass defect & $\DeltaM \bar m_0 c^2 \lambda_j(t)/2$ & & & $\phantom{-} 0$ \\  \cline{2-6}
    & oscillation of transition energy & $\DeltaMDM(t) \bar m_0 c^2 \lambda_j(t)/2$ & $\phi_m$ & $=$ & $- \varrho_0 (\Delta \varepsilon \omega_c + \bar \varepsilon \Omega) \tau_1(t_0)$ \\ \hline
    \multirow{6}{*}{$\hat{V}_\mathrm{kin}$} & \multirow{2}{*}{mean-mass oscillation} & \multirow{2}{*}{$- \MeanMDM(t) \frac{\hat{p}^2}{2 \bar{m}_0}$} & $\phi_1$ & $=$ & $\phantom{-} \bar{\varepsilon} \varrho_0 \omega_k \tau_1(t_0) (1 + 2 \wp_0)$ \\
    & & & $\phi_2$ & $=$ & $- \bar{\varepsilon} \varrho_0 k g_0 \tau_2^2(t_0)$ \\  \cline{2-6}
    & mass defect & $- \DeltaM \frac{\hat{p}^2}{2 \bar{m}_0} \lambda_j(t)/2$ & $\phi_3$ & $=$ & $\phantom{-} \DeltaM k g_0 T^2 \bar{\wp}_{t_0+T}$ \\  \cline{2-6}
    & \multirow{3}{*}{oscillation of transition energy} & \multirow{3}{*}{$- \DeltaMDM(t) \frac{\hat{p}^2}{2 \bar{m}_0} \lambda_j(t)/2$} & $\phi_4$ & $=$ & $\phantom{-} \Delta \varepsilon \varrho_0 \omega_k \tau_1(t_0) (\wp_0^2 + \wp_0 + 1/2)$ \\
    & & & $\phi_5$ & $=$ & $- \Delta \varepsilon \varrho_0 k g_0 \tau_2^2(t_0) (\wp_0 + 1/2)$ \\
    & & & $\phi_6$ & $=$ & $\phantom{-} \Delta \varepsilon \varrho_0 \frac{g_0^2}{2 c^2} \omega_c \tau_3^3(t_0)$ \\ \hline
    \multirow{8}{*}{$\hat{V}_\mathrm{pot}$} & EP violation & $\MeanGEP \bar{m}_0 g_0 \hat{z}$ & $\phi_7$ & $=$ & $- \varepsilon_S \bar \varepsilon k g_0 T^2$ \\ \cline{2-6}
    & state-dependent EP violation & $\DeltaGEP \bar{m}_0 g_0 \hat{z} \lambda_j(t)/2$ & $\phi_8$ & $=$ & $\phantom{-} \varepsilon_S \Delta \varepsilon k g_0 T^2 \bar{\wp}_{t_0+T}$ \\  \cline{2-6}
    & oscillation of gravity & $\MeanGDM(t) \bar{m}_0 g_0 \hat{z}$ & $\phi_9$ & $=$ & $- \varepsilon_S \varrho_0 k g_0 \tau_S^2(t_0)$ \\ \cline{2-6}
    & oscillation of transition energy & $\MeanMDM(t) \bar{m}_0 g_0 \hat{z}$ & $\phi_{10}$ & $=$ & $- \bar \varepsilon \varrho_0 k g_0 \tau_\mathrm{EP}^2(t_0)$ \\  \cline{2-6}
    & mass defect & $\DeltaM \bar{m}_0 g_0 \hat{z} \lambda_j(t)/2$ & $\phi_{11}$ & $=$ & $\phantom{-} \DeltaM k g_0 T^2 \bar{\wp}_{t_0+T}$ \\  \cline{2-6}
    & \multirow{3}{*}{oscillation of transition energy} & \multirow{3}{*}{$\DeltaMDM(t) \bar{m}_0 g_0 \hat{z} \lambda_j(t)/2$} & $\phi_{12}$ & $=$ & $-\Delta \varepsilon \varrho_0 \frac{g_0 z_0}{c^2} \omega_c \tau_1(t_0)$ \\
    & & & $\phi_{13}$ & $=$ & $-\Delta \varepsilon \varrho_0 k g_0 \tau_2^2(t_0)(\wp_0 + 1/2)$ \\
    & & & $\phi_{14}$ & $=$ & $\phantom{-} \Delta \varepsilon \varrho_0 \frac{g_0^2}{2 c^2} \omega_c \tau_3^3(t_0)$
\end{tabular}
\end{ruledtabular}
\end{table*}


\begin{table*}[h!]
\caption{%
\label{tab:time_scales}%
Time scales relevant for the individual phase contributions listed in Table~\ref{tab:mzi_phases} that naturally arise in the perturbative calculation.~\cite{Ufrecht2020} %
They depend on the initial time $t_0$ and are obtained by averaging of the phase $\vartheta=\omega_\varrho t +\phi_\varrho$. %
For a compact notation we defined $\langle h \rangle \big\vert^{t_2}_{t_1} \coloneqq \int_{t_1}^{t_2} \! \text{d}t \, h(t)$.%
}
\renewcommand{\arraystretch}{2}
\begin{ruledtabular}
\begin{tabular}{rl}
Time Scale & Definition \\ \hline
$\tau_1(t_0)$ & $\langle \cos \vartheta \rangle \big\vert^{t_0+T}_{t_0} - \langle \cos \vartheta \rangle \big\vert^{t_0+2T}_{t_0+T}$ \\ \hline
$\tau_2^2(t_0)$ & $\langle (t-t_0) \cos \vartheta \rangle \big\vert^{t_0+T}_{t_0} - \langle (t-t_0) \cos \vartheta \rangle \big\vert^{t_0+2T}_{t_0+T}$ \\ \hline
$\tau_3^3(t_0)$ & $\langle (t-t_0)^2 \cos \vartheta \rangle \big\vert^{t_0+T}_{t_0} - \langle (t-t_0)^2\cos \vartheta \rangle \big\vert^{t_0+2T}_{t_0+T}$ \\ \hline
$\tau_S^2(t_0)$ & $\langle (t-t_0) \cos(\vartheta+\phi_S) \rangle \big\vert^{t_0+T}_{t_0} + \langle [2T - (t-t_0)] \cos(\vartheta+\phi_S) \rangle \big\vert^{t_0+2T}_{t_0+T}$ \\ \hline
$\tau_\mathrm{EP}^2(t_0)$ & $\langle (t-t_0) \cos \vartheta \rangle \big\vert^{t_0+T}_{t_0} + \langle [2T-(t-t_0)]\cos \vartheta \rangle \big\vert^{t_0+2T}_{t_0+T}$
\end{tabular}
\end{ruledtabular}
\end{table*}

\begin{table*}[h!]
\caption{\label{tab:sp_signal_contributions}%
Non-vanishing and next sub-leading contributions $\big\langle \Delta \phi^2_{m,j} \big \rangle$ to the signal amplitude for single-photon Mach-Zehnder interferometers. %
The signal contributions are calculated by averaging the square of a differential signal in accordance with Eq.~\eqref{eq:signal_contrib} and based on the phases provided in Table~\ref{tab:mzi_phases}. %
The dominant contribution, \ie{} $\big\langle \Delta \phi^2_{m,m} \big \rangle$ is given in the main body of the article. %
The explicit expressions of these terms are split into a numerical factor, a scale, the violation parameters, and some interferometric factor that includes the interrogation-mode function.~\cite{dipumpo2023optimal}
For different interferometer geometries, multiple loops, or large momentum transfer one expects other interferometric and numerical factors. %
Note that we have $\big\langle \Delta \phi^2_{m,13} \big \rangle = \big\langle \Delta \phi^2_{m,5} \big \rangle$ and $\big\langle \Delta \phi^2_{m,14} \big \rangle = \big\langle \Delta \phi^2_{m,6} \big \rangle$. %
We also introduced $\mathcal{C}(t) \coloneqq \cos \!\left(\omDM t/2\right)$.%
}
\renewcommand{\arraystretch}{2.25}
\begin{ruledtabular}
\begin{tabular}{lcrcccccr}
& & Factor & & Scale & & Violation Parameters & & Interferometric Factor\\[0.5em] \hline
$\big\langle \Delta \phi^2_{m,1} \big \rangle/\frac{\Delta \varepsilon \omega_c + \bar{\varepsilon} \Omega}{\omDM}$ &%
$=$ &%
$-16$ &%
$\times$ &%
$\displaystyle \frac{\omega_k}{\omDM}$ &%
$\times$ &%
$\varrho_0^2 \bar \varepsilon$ &%
$\times$ &%
$( 1 + 4 \bar \wp_0 ) \SigSin{\tau_L}{2} \SigSin{T}{4}$ \\[0.5em] \hline
$\big\langle \Delta \phi^2_{m,2} \big \rangle/\frac{\Delta \varepsilon \omega_c + \bar{\varepsilon} \Omega}{\omDM}$ &%
$=$ &%
$32$ &%
$\times$ &%
$\displaystyle \frac{k g_0 T}{\omDM}$ &%
$\times$ &%
$\varrho_0^2 \bar \varepsilon$ &%
$\times$ &%
$\SigSin{\tau_L}{2} \SigSin{T}{4}$ \\[0.5em] \hline
$\big\langle \Delta \phi^2_{m,4} \big \rangle/\frac{\Delta \varepsilon \omega_c + \bar{\varepsilon} \Omega}{\omDM}$ &%
$=$ &%
$-8$ &%
$\times$ &%
$\displaystyle \frac{\omega_k}{\omDM}$ &%
$\times$ &%
$\varrho_0^2 \Delta \varepsilon$ &%
$\times$ &%
$ \left[ 2 + 4 \bar \wp_0 (1 + \bar \wp_0) + \Delta \wp_0^2 \right] \SigSin{\tau_L}{2} \SigSin{T}{4}$ \\[0.5em] \hline
$\big\langle \Delta \phi^2_{m,5} \big \rangle/\frac{\Delta \varepsilon \omega_c + \bar{\varepsilon} \Omega}{\omDM}$ &%
$=$ &%
$4$ &%
$\times$ &%
$\displaystyle \frac{k g_0}{\omDM^2}$ &%
$\times$ &%
$\varrho_0^2 \Delta \varepsilon$ &%
$\times$ &%
{$\begin{aligned}
    &\SigSin{T}{2} \bigg\{ 2 (1 + 2 \bar \wp_0) \omDM T \left[ \SigCos{2\tau_L}{1} - 1 \right]\SigSin{T}{2} \\
    &+ \Delta \wp_0 \SigSin{2\tau_L}{1} \left[ \SigCos{2T}{1} + \omDM T \SigSin{2T}{1} - 1 \right] \bigg\}
\end{aligned}$} \\[0.5em] \hline
$\big\langle \Delta \phi^2_{m,6} \big \rangle/\frac{\Delta \varepsilon \omega_c + \bar{\varepsilon} \Omega}{\omDM}$ &%
$=$ &%
$8$ &%
$\times$ &%
$\displaystyle \frac{g_0^2}{c^2 \omDM^2} \frac{\omega_c}{\omDM}$ &%
$\times$ &%
$\varrho_0^2 \Delta \varepsilon$& %
$\times$ &%
{$\begin{aligned}
    \textstyle \SigSin{\tau_L}{2} \SigSin{T}{2} &\bigg\{ 2 + 2( \omDM^2 T^2 - 1 )\SigCos{2T}{1} \\
    &\textstyle - \omDM^2 T^2 - 2 \omDM T \SigSin{2T}{1} \bigg\}
\end{aligned}$} \\[0.5em] \hline
$\big\langle \Delta \phi^2_{m,9} \big \rangle/\frac{\Delta \varepsilon \omega_c + \bar{\varepsilon} \Omega}{\omDM}$ &%
$=$ &%
$-32$ &%
$\times$ &%
$\displaystyle \frac{k g_0}{\omDM^2}$ &%
$\times$ &%
$\varrho_0^2 \varepsilon_S$ &%
$\times$ &%
$\sin \phi_S \SigSin{\tau_L}{2} \SigSin{T}{4}$\\[0.5em] \hline
$\big\langle \Delta \phi^2_{m,12} \big \rangle/\frac{\Delta \varepsilon \omega_c + \bar{\varepsilon} \Omega}{\omDM}$ &%
$=$ &%
$16$ &%
$\times$ &%
$\displaystyle \frac{g_0 (2 z_0 + L)}{c^2} \frac{\omega_c}{\omDM}$ &%
$\times$ &%
$\varrho_0^2 \Delta \varepsilon$ &%
$\times$ &%
$\SigSin{\tau_L}{2} \SigSin{T}{4}$\\
\end{tabular}
\end{ruledtabular}
\end{table*}

\begin{table*}[h!]
\caption{%
\label{tab:bragg_signal_contributions}%
Non-vanishing contributions $\big\langle \Delta \phi^2_{i,j} \big \rangle$ to the signal amplitude for Bragg-type Mach-Zehnder interferometers. %
The signal contributions are calculated by averaging the square of a differential signal in accordance with Eq.~\eqref{eq:signal_contrib} and based on the phases provided in Table~\ref{tab:mzi_phases}.
Their explicit expression is split into a numerical factor, a scale, the violation parameters, and some interferometric factor. For different interferometers one expects other interferometric and numerical factors. %
In contrast to single-photon transitions, no dominant scale can be identified.
}
\renewcommand{\arraystretch}{2.5}
\begin{ruledtabular}
\begin{tabular}{lcrcccccr}
& & Factor & & Scale & & Violation Parameters & & Interferometric Factor\\[0.5em]
\hline
$\big\langle \Delta \phi^2_{9,9} \big \rangle$ &%
$=$ &%
$32$ &%
$\times$ &%
$\displaystyle \frac{k^2 g_0^2}{\omDM^4}$ &%
$\times$ &%
$\varrho_0^2 \varepsilon_S^2$ &%
$\times$ &%
$\SigSin{\tau_L}{2} \SigSin{T}{4}$ \\[0.5em] \hline
$\big\langle \Delta \phi^2_{9,10} \big \rangle$ &%
$=$ &%
$32$ &%
$\times$ &%
$\displaystyle \frac{k^2 g_0^2}{\omDM^4}$ &%
$\times$ &%
$\varrho_0^2 \bar \varepsilon \varepsilon_S$ &%
$\times$ &%
$\cos \phi_S \SigSin{\tau_L}{2} \SigSin{T}{4}$ \\[0.5em] \hline
$\big\langle \Delta \phi^2_{9,1} \big \rangle$ &%
$=$ &%
$- 16$ &%
$\times$ &%
$\displaystyle \frac{k g_0}{\omDM^2} \frac{\omega_k}{\omDM}$ &%
$\times$ &%
$\varrho_0^2 \bar \varepsilon \varepsilon_S$ &%
$\times$ &%
$\SigSin{T}{4} \left[ \Delta \wp_0 \cos \phi_S \SigSin{2\tau_L}{1} - 2 (1+2 \bar \wp_0) \sin \phi_S \SigSin{\tau_L}{2} \right]$ \\[0.5em] \hline
$\big\langle \Delta \phi^2_{9,2} \big \rangle$ &%
$=$ &%
$32$ &%
$\times$ &%
$\displaystyle \frac{k^2 g_0^2}{\omDM^4}$ &%
$\times$ &%
$\varrho_0^2 \bar \varepsilon \varepsilon_S$ &%
$\times$ &%
$\SigSin{\tau_L}{2} \SigSin{T}{4} \left\{ \cos \phi_S - \omDM T \left[ \sin \phi_S + \SigCos{T}{1}/\SigSin{T}{1} \cos \phi_S \right] \right\}$ \\[0.5em] \hline
$\big\langle \Delta \phi^2_{10,10} \big \rangle$ &%
$=$ &%
$32$ &%
$\times$ &%
$\displaystyle \frac{k^2 g_0^2}{\omDM^4}$ &%
$\times$ &%
$\varrho_0^2 {\bar\varepsilon}^2$ &%
$\times$ &%
$\SigSin{\tau_L}{2} \SigSin{T}{4}$ \\[0.5em] \hline
$\big\langle \Delta \phi^2_{10,1} \big \rangle$ &%
$=$ &%
$- 16$ &%
$\times$ &%
$\displaystyle \frac{k g_0}{\omDM^2} \frac{\omega_k}{\omDM}$ &%
$\times$ &%
$\varrho_0^2 {\bar\varepsilon}^2$ &%
$\times$ &%
$\Delta \wp_0 \SigSin{2\tau_L}{1} \SigSin{T}{4}$ \\[0.5em] \hline
$\big\langle \Delta \phi^2_{10,2} \big \rangle$ &%
$=$ &%
$32$ &%
$\times$ &%
$\displaystyle \frac{k^2 g_0^2}{\omDM^4}$ &%
$\times$ &%
$\varrho_0^2 {\bar\varepsilon}^2$ &%
$\times$ &%
$\SigSin{\tau_L}{2} \SigSin{T}{3} \left[ \SigSin{T}{1} - \omDM T \SigCos{T}{1} \right]$ \\[0.5em] \hline
$\big\langle \Delta \phi^2_{1,1} \big \rangle$ &%
$=$ &%
$16$ &%
$\times$ &%
$\displaystyle \frac{\omega_k^2}{\omDM^2}$ &%
$\times$ &%
$\varrho_0^2 {\bar\varepsilon}^2$ &%
$\times$ &%
$\SigSin{T}{4} \left\{ \left[ 1 - \SigCos{2\tau_L}{1} \right] \left[ 1 + 4 \bar \wp_0 (1 + \bar \wp_0)\right] + \Delta \wp_0 \left[ 1 + \SigCos{2\tau_L}{1} \right] \right\}$ \\[0.5em] \hline
$\big\langle \Delta \phi^2_{1,2} \big \rangle$ &%
$=$ &%
$- 8$ &%
$\times$ &%
$\displaystyle \frac{k g_0}{\omDM^2} \frac{\omega_k}{\omDM}$ &%
$\times$ &%
$\varrho_0^2 {\bar\varepsilon}^2$ &%
$\times$ &%
{$\begin{aligned}
    &\SigSin{T}{2} \Big\{4 (1 + 2 \bar \wp _0) \omDM T \SigSin{\tau_L}{2} \SigSin{T}{2} \\
    &+ \Delta \wp_0 \SigSin{2\tau_L}{1}[1 - \SigCos{2T}{1} - \omDM T \SigSin{2T}{1}] \Big\}
\end{aligned}$} \\[0.5em] \hline
$\big\langle \Delta \phi^2_{2,2} \big \rangle$ &%
$=$ &%
$16$ &%
$\times$ &%
$\displaystyle \frac{k^2 g_0^2}{\omDM^4}$ &%
$\times$ &%
$\varrho_0^2 {\bar\varepsilon}^2$ &%
$\times$ &%
$\SigSin{\tau_L}{2} \SigSin{T}{2} \left[ 1 + 2 \omDM^2 T^2 - \SigCos{2T}{1} - 2 \omDM T \SigSin{2T}{1} \right]$
\end{tabular}
\end{ruledtabular}
\end{table*}

\clearpage

\section*{References}
\bibliography{references}

\end{document}